# A Novel Temperature-Induced Magnetization Transition in Pt/Co/Pt Sandwiches

W. W. Lin (林维维) and H. Sang (桑海)

*National Laboratory of Solid State Microstructures and Department of Physics,*
*Nanjing University, Nanjing, Jiangsu 210093, China*

(Submitted 12 December 2008)

It is found in the Pt/Co/Pt(111) sandwiches that perpendicular magnetic anisotropy weakens associated with the dramatic magnetization enhancement as the temperature decreases below a critical temperature, $T_{\text{cri}}$, which increases with the Pt overlayer thickness $t^{\text{Pt}}$. For the sandwich with enough thick Pt overlayer, it trends to an unusual state with weak perpendicular magnetic anisotropy in the absence of obvious easy direction behavior at low temperature. It suggests that Pt is more significantly polarized in the film with thicker Pt overlayer. The results reveal the strong influence of both thick nonmagnetic coverage and temperature on the electronic structure of the Pt/Co/Pt(111) sandwich.

Ferromagnetic (FM)/nonmagnetic (NM) metal films have been extensively studied in the last two decades due to the interesting physics and the wide applications in the magnetic storage media and spintronic devices [1−25]. The magnetization and magnetic anisotropy of the FM layer in the FM/NM film depend on not only the FM layer itself [1−5] but also the NM layer [6−20]. In this case, the NM metal is spin polarized due to the electron hybridization at the FM/NM interface [21−25]. The NM coating can increase the Cuire temperature of FM monolayer [6,7]. The magnetization, the remanent ratio, and the coercive field of the FM/NM film show nonmonotonic variation with the NM overlayer thickness [8−10]. With the NM (e.g. Cu, Pd, Ag, and Au) overlayer on the Co/Pd(111) [8] and Co/Au(111) [9], perpendicular magnetic anisotropy (PMA) is enhanced rapidly with a maximum at about one monolayer coverage, which can be well explained in terms of overlayer induced changes in the electronic structure [11]. In Cu/Co/Cu(001) film, the magnetic anisotropy field oscillates with the thickness of the Cu overlayer which is related to the occurrence of quantum well (QW) states [12,13]. In Au/Co/Au(111) films, the spin reorientation transition (SRT) from out-of-plane to in-plane takes place when increasing the thickness of the Au cap layer, which is linked with the decrease of the Co orbital moment [18]. However, most experimental and theoretical researches on the NM overlayer effects focus on the properties in the case of the coverage within several monolayers, less in that of the thicker coverage.

Besides the NM overlayer, the magnetic properties are also influenced by the temperature. Usually, the magnetization, perpendicular magnetic anisotropy, and coercive field increase with temperature decreasing [20]. The SRT from in-plane to out-of-plane may happen as temperature decreases in the systems such as Fe/Cu and Co/Au films [1,15]. In Ni/Cu(001) film, however, the easy magnetization direction changes from out-of-plane to in-plane with temperature decreasing, which is explained by the difference in the temperature dependence of the surface and volume anisotropy energy [14]. Both temperature and thickness induced reorientation of magnetization may arise from the changes in the electronic structure and could be microscopically investigated with the help of spin- and layer- dependent density of states (DOS) [16].

In this Letter, we report on a novel temperature-induced magnetization transition in Pt/Co/Pt(111) sandwiches, which has not ever been observed. Pt/Co/Pt(111) film is known as a typical PMA system with an easy axis perpendicular to the film plane. However, it is found that PMA *weakens* as the temperature *decreases* below a critical temperature, $T_{cri}$. For the sandwich with enough thick Pt overlayer, it trends to an unusual state with weak PMA in the absence of obvious easy direction behavior, which is different from the usual SRT where the easy magnetization direction changes from in-plane to out-of-plane (or from out-of-plane to in-plane). The transition depends on the Pt overlayer thickness. Our study was carried out in a very wide range of Pt overlayer thickness up to 1000 Å. The novel behavior is more pronounced in the film with thick Pt coverage. The magnetization of the films with thick Pt overlayer enhances dramatically which probably imply a new ferromagnetic phase appearing as the temperature decreases below $T_{cri}$. The present result reveals an exciting and basic fact in the magnetic film, and is important to push forward the fundamental understanding to the magnetization and magnetic anisotropy of the FM/NM film.

The Pt/Co/Pt sandwiches were deposited on the SiO$_2$/Si substrates at room temperature (RT) using a dc magnetron sputtering system. The base pressure of the sputtering system was $1.2 \times 10^{-8}$ Torr, and the Ar pressure during sputtering was about 4 mTorr. A Pt seed layer of 55 Å thickness was grown first on the substrate to achieve (111) texture, and then a 5 Å thick Co layer was deposited followed by a Pt overlayer. The depositing rates of Co and Pt were 0.22 Å/s and 0.35 Å/s, respectively. The sandwiches with various Pt overlayer thicknesses $t^{Pt}$ from 10 Å to 1000 Å were prepared. X-ray diffraction indicated preferential (111) texture in the film.

The perpendicular and in-plane magnetic hysteresis loops of the samples were measured using a superconducting quantum interference device (SQUID) with an applied magnetic field $H$ up to 20 kOe and the temperature $T$ from 5 K to 300 K. All Pt/Co/Pt sandwiches exhibit square loops at RT when the field is applied perpendicular to the film plane, indicating perpendicular easy axis at RT.

Figure 1(a)−(c) show the perpendicular and in-plane hysteresis loops of the Pt/Co/Pt sandwich with a very small Pt overlayer thickness $t^{Pt}$ = 10 Å at $T$ = 5, 180, and 300 K, respectively. The hysteresis loops are square with the remanent ratio $M_r^\perp/M_S \sim 1$ for the applied field perpendicular to the film plane, but tilted for the in-plane field. It indicates that a perpendicular easy axis in the Pt(10 Å)/Co/Pt sandwich remains in the range of $T$ from 5 K to 300 K. Both the saturation magnetization and the perpendicular coercive field of the film increase with the decrease of $T$.

The Pt/Co/Pt sandwiches with thick Pt overlayer, however, display unusual features. Shown in Fig. 1(d)−(g) are the perpendicular and in-plane hysteresis loops of the Pt/Co/Pt sandwich with a large Pt overlayer thickness $t^{Pt}$ = 500 Å. At $T$ = 300 K [Fig. 1(d)] and 180 K [Fig. 1(e)], the Pt(500 Å)/Co/Pt sandwich exhibit square perpendicular hysteresis loop and tilted in-plane hysteresis loop, indicating the easy axis is still perpendicular to the film plane. While at $T$ = 120 K [Fig. 1(f)], the perpendicular hysteresis loop of the Pt(500 Å)/Co/Pt sandwich is no longer square and its $M_r^\perp/M_S$ is about 0.53 only. At $T$ = 5 K, the saturation magnetization enhances significantly, and the perpendicular hysteresis loop is tilted and close to the in-plane one [Fig. 1(g)]. One can see that the perpendicular hysteresis loop changes from square to slant and trends to close to the in-plane one as $T$ decreases below 120 K, indicating the weakening of PMA and the absence of obvious easy direction behavior. It suggests that a transition of the magnetization reversal dynamics take place as $T$ reducing.

Figure 2(a) shows the dependence of the perpendicular remanent ratio $M_r^\perp/M_S$ on $T$ for various Pt overlayer thicknesses. For $t^{Pt} = 50$ Å, as $T$ reducing, $M_r^\perp/M_S$ remains 1, but decreases below a critical temperature $T_{cri} \sim 50$ K, and then increases below 20 K. For $t^{Pt} = 500$ Å, $M_r^\perp/M_S$ drops largely with $T$ decreasing below $T_{cri} \sim 130$ K, and has a minimum of about 0.07 at $T \sim 70$ K. The decrease of $M_r^\perp/M_S$ below $T_{cri}$ means that the easy magnetization direction deviates from the normal of the film plane. The increase of $M_r^\perp/M_S$ with $T$ decreasing is due to the broadening of perpendicular hysteresis loop with the enhanced coercive field. One can see from Fig. 2(a) that $T_{cri}$ increases, while the minimum of $M_r^\perp/M_S$ decreases with $t^{Pt}$ increasing.

The critical temperature $T_{cri}$ increases fast with the Pt overlayer thickness $t^{Pt} < 200$ Å, but slowly with $t^{Pt} > 200$ Å, as shown in the inset of Fig. 2(a). We find that $T_{cri}$ fits well with a logarithmic function of $t^{Pt}$ expressed as

$$T_{cri}(t^{Pt}) = a\ln(t^{Pt} - t_0) - T_0, \tag{1}$$

where $a$, $t_0$, and $T_0$ are parameters for the fitting. In our cases, $a = 36.9$, $t_0 = 20$ Å, and $T_0 = 78.2$ K. From Eq. (1), one can see that the transition takes place above zero temperature only when $t^{Pt}$ is larger than a certain thickness. In our cases, the transition could take place when $t^{Pt} > 28.3$ Å.

Shown in Fig. 2(b) is the dependence of the perpendicular coercive field $H_C^\perp$ on $T$ for the films with various Pt overlayer thicknesses. For $t^{Pt} = 10$ Å, $H_C^\perp$ increases monotonically with the decrease of $T$, seeing the inset of Fig. 2(b), which is usually observed in the FM materials [20]. However, the unusual behavior of $H_C^\perp$ is observed in the films with thick Pt overlayer. For $t^{Pt} = 100$ Å, $H_C^\perp$ varies flatly for $T$ between 50 K and 100 K. For $t^{Pt} = 500$ Å, as $T$ decreases, $H_C^\perp$ increases from 300 K to 140 K, but decreases from 140 K to 80 K, and then increases again below 80 K. It is noted that the decrease of $H_C^\perp$ occurs near to $T_{cri}$ and is related to the change of magnetic anisotropy.

Figure 3 shows the temperature dependence of the magnetization of the films for various Pt overlayer thicknesses. The magnetization of the film with 100 Å Pt overlayer enhances about 70% as $T$ decreases from 300 K to 5 K. The magnetization of the films with 200 Å and 500 Å Pt overlayer enhance largely, about 126% and 150%, respectively, as $T$ decreases from 300 K to 5 K. Significant enhancement of the magnetization of the films with 200 Å and 500 Å Pt overlayer occur around 90 K and 120 K, respectively, which is close to $T_{cri}$. It suggests that it seems a new ferromagnetic phase below $T_{cri}$. However, it is still an open question.

Actually, the measured magnetization is contributed by both Co and Pt, which can be written as

$$M_S t^{Co} = M_S^{Co} t^{Co} + M_S^{Pt} t_{eff}^{Pt}, \qquad (2)$$

where $M_S$ is the total saturation magnetization averaging to Co, $M_S^{Co}$ and $t^{Co}$ are the saturation magnetization and the thickness of Co layer, respectively, and $M_S^{Pt}$ and $t_{eff}^{Pt}$ are the average saturation magnetization and the effective thickness of polarized Pt layer, respectively. The total magnetization averaging to Co is given by $M_S = M_S^{Co} + M_S^{Pt} t_S^{Pt}/t^{Co}$ from Eq. (2). The Co moment is as large as 1.69 $\mu_B$/atom (about 1700 emu/cm$^3$) in the Co/Pt multilayer at RT [23]. The maximum magnetic moment for Co by the Hund rule is 3 $\mu_B$/atom. The Pt layer is polarized due to the Co 3d-Pt 5d hybridization at the interface [21−25]. In the Co/Pt multilayer at RT, the polarized Pt moment is 0.21 $\mu_B$/atom (about 210 emu/cm$^3$) at the Pt/Co interface, followed by a decay of the induced polarization within 1 nm [25]. For the sandwich with $t^{Pt}$ = 500 Å, $M_S$ reaches up to about 4800 emu/cm$^3$ at $T$ = 5 K, which is much larger than $M_S^{Co}$. It suggests that the polarized Pt magnetization may enhance and/or the effective thickness $t_{eff}^{Pt}$ may increase below $T_{cri}$, besides the Co magnetization involving spin and orbital moments enhances at low temperature. This effect is pronounced particularly in the sandwiches with thicker Pt overlayer.

The magnetic anisotropy of the film is strongly related to the magnetization. For the Pt/Co/Pt film, the magnetic anisotropies that need to be considered include the interfacial magnetocrystalline anisotropy resulting from the spin-orbit interaction and the shape anisotropy arising from the magnetostatic dipole-dipole interaction. The magnetic anisotropy of Pt, which was not included due to its insignificant contribution in usual situations, should be taken into account owing to considerable polarization of Pt. The effective anisotropy energy $K_{eff} t^{Co}$ can be written as

$$K_{eff} t^{Co} = K_S^{Co} + K_S^{Pt} + \left(K_V^{Co} + K_D^{Co}\right) t^{Co} + \left(K_V^{Pt} + K_D^{Pt}\right) t_{eff}^{Pt}. \qquad (3)$$

$K_S^{Co}$ is the interface magnetic anisotropy of the Co layer including asymmetric contributions of the interface to the Pt sublayer and the one to the Pt overlayer. The interfacial spin-orbit coupling which is related to the anisotropy of the orbital magnetic moment at the interface between Co and Pt layers makes the Co magnetization favor the out-of-plane orientation. $K_S^{Pt}$ is the interface magnetic anisotropy of the Pt layer, which may make the Pt magnetization prefer in plane. $K_V^{Co}$ and $K_D^{Co}$ are the volume anisotropy energy and the shape anisotropy of the Co layer, respectively, and $K_V^{Pt}$ and $K_D^{Pt}$ are those of the Pt layer. The dipole-dipole coupling makes the magnetization prefer lying in plane. The effective anisotropy field $H_K = 2K_{eff}/M_S$ is given by

$$H_K = \frac{2}{M_S}\left[\frac{K_S^{Co} + K_S^{Pt}}{t^{Co}} + K_V^{Co} + K_D^{Co} + \left(K_V^{Pt} + K_D^{Pt}\right)\frac{t_{eff}^{Pt}}{t^{Co}}\right]. \quad (4)$$

The contribution of the interface magnetic anisotropy is proportional to the inverse of the Co layer thickness, as shown in Eq. (4). When the Co layer is very thin, the Pt/Co/Pt sandwich has a perpendicular easy axis due to that the interface magnetic anisotropy exceeds the shape anisotropy. This is the behavior above $T_{cri}$. However, the results show that PMA weakens below $T_{cri}$. It suggests that the decrease of the effective anisotropy field $H_K$ below $T_{cri}$ is related to the magnetization enhancement of the film with the temperature decreasing. The shape anisotropy increases due to the enhancement of the Co and Pt magnetization. The increasing of both the magnetization and the effective thickness of the polarized Pt increases the contribution of the magnetic anisotropy of Pt. These weaken the contribution of the interface magnetic anisotropy of Co, and thus yield the weakening of PMA.

The magnetization and the magnetic anisotropy are strongly related to the electronic structure of the film, in particular, density of states. The magnetization enhancement of the film below $T_{cri}$ reveals the strong influence of temperature on the electronic structure of the film. $T_{cri}$ increases with $t^{Pt}$, indicating that the Pt layer can be more easily and strongly polarized for the sandwiches with thicker Pt coverage. In other words, the thickness of Pt coverage can influence the temperature dependence of the electronic structure of the film.

In summary, a novel temperature-induced magnetization transition has been found in Pt/Co/Pt(111) sandwiches. PMA weakens as the temperature decreases below a $T_{cri}$ which increases with $t^{Pt}$. For the sandwich with enough thick Pt overlayer, it trends to an unusual state with weak PMA in the absence of obvious easy direction behavior at low temperature, which is different from the usual SRT. The magnetization of the films with thick Pt overlayer enhances dramatically which probably imply a new ferromagnetic phase appearing as the temperature decreases below $T_{cri}$. It suggests that Pt is more significantly polarized in the film with thicker Pt overlayer. The weakening of PMA below $T_{cri}$ is closely related to the dramatic magnetization enhancement of the film. The magnetic anisotropy of Pt, which was not included due to its insignificant contribution in usual situations, should be taken into account owing to considerable polarization of Pt. The results reveal the strong influence of the thick

NM coverage and temperature on the electronic structure of the sandwich, which may open a new sight to understand the magnetic behavior in the FM/NM film.

Further study to make clear the Co and Pt contributions to the magnetization enhancement below $T_{cri}$, the experimental observation using the element-specific and spin-dependent spectroscopy at low temperature is a good way, e.g. x-ray magnetic circular dichroism (XMCD) spectroscopy. Moreover, the magnetic microscopy could be employed to observe the change of domain configuration, in particular, near to $T_{cri}$ where a transition of the magnetization reversal dynamics may take place. Besides the experimental exploring of the origin of the novel transition, theoretical consideration about the influence of both thick NM coverage and temperature on the electronic structure of the film is expected.

This work was partly supported by NSFC Grant Nos. 10574065 and 10128409, NBRPC Grant No. 2009CB929503, and JSNSF.

References

[1] R. Allenspach and A. Bischof, Phys. Rev. Lett. **69**, 3385 (1992).

[2] F. Huang, M. T. Kief, G. J. Mankey, and R. F. Willis, Phys. Rev. B **49**, 3962 (1994).

[3] J. Dorantes-Dávila, H. Dreyssé, and G. M. Pastor, Phys. Rev. Lett. **91**, 197206 (2003).

[4] F. El Gabaly, S. Gallego, C. Munoz, L. Szunyogh, P. Weinberger, C. Klein, A. K. Schmid, K. F. McCarty, and J. de la Figuera, Phys. Rev. Lett. **96**, 147202 (2006).

[5] J. W. Lee, J. Kim, S. K. Kim, J. R. Jeong, and S. C. Shin, Phys. Rev. B **65**, 144437 (2002).

[6] M. Przybylski and U. Gradmann, Phys. Rev. Lett. **59**, 1152 (1987).

[7] W. Weber, D. Kerkmann, D. Pescia, D. A. Wesner, and G. Güntherodt, Phys. Rev. Lett. **65**, 2058 (1990).

[8] B. N. Engel, M. H. Wiedmann, R. A. Van Leeuwen, and C. M. Falco, Phys. Rev. B **48**, 9894 (1993).

[9] P. Beauvillain, A. Bounouh, C. Chappert, S. Ould-Mahfoud, J. P. Renard, P. Veillet, D. Weller, and J. Corno, J. Appl. Phys. **76**, 6078 (1994).

[10] M. E. Buckley, F. O. Schumann, and J. A. C. Bland, Phys. Rev. B **52**, 6596 (1995).

[11] B. Újfalussy, L. Szunyogh, P. Bruno, and P. Weinberger, Phys. Rev. Lett. **77**, 1805 (1996).

[12] W. Weber, A. Bischof, R. Allenspach, C. Würsch, C. H. Back, and D. Pescia, Phys. Rev. Lett. **76**, 3424 (1996).

[13] C. Würsch, C. Stamm, S. Egger, D. Pescia, W. Baltensperger, and J. S. Helman, Nature **389**, 937 (1997).

[14] M. Farle, W. Platow, A. N. Anisimov, P. Poulopoulos, and K. Baberschke. Phys. Rev. B **56**, 5100 (1997).

[15] R. Sellmann, H. Fritzsche, H. Maletta, V. Leiner, and R. Siebrecht, Phys. Rev. B **64**, 054418 (2001).

[16] J. H. Wu, H. Y. Chen, and W. Nolting. Phys. Rev. B **65**, 014424 (2001).

[17] M. Kisielewski, A. Maziewski, M. Tekielak, A. Wawro, and L. T. Baczewski, Phys. Rev. Lett. **89**, 087203 (2002).

[18] J. Langer, J. H. Dunn, A. Hahlin, O. Karis, R. Sellmann, D. Arvanitis, and H. Maletta, Phys. Rev. B **66**, 172401 (2002).

[19] A. Enders, D. Peterka, D. Repetto, N. Lin, A. Dmitriev, and K. Kern, Phys. Rev. Lett. **90**, 217203 (2003).

[20] T. Suzuki, H. Notarys, D. C. Dobbertin, C. J. Lin, D. Weller, D. C. Miller, and G.

Gorman, IEEE Trans. Magn. 28, 2754 (1992).

[21] G. Schütz, R. Wienke, W. Wilhelm, W. B. Zeper, H. Ebert, and K. Spörl, J. Appl. Phys. **67**, 4456 (1990).

[22] S. Rüegg, G. Schütz, P. Fischer, R. Wienke, W. B. Zeper, and H. Ebert, J. Appl. Phys. **69**, 5655 (1991).

[23] J. Thiele, C. Boeglin, K. Hricovini, and F. Chevrier, Phys. Rev. B **53**, R11934 (1996).

[24] N. Nakajima, T. Koide, T. Shidara, H. Miyauchi, H. Fukutani, A. Fujimori, K. Iio, T. Katayama, M. Nyvlt, and Y. Suzuki, Phys. Rev. Lett. **81**, 5229 (1998).

[25] J. Geissler, E. Goering, M. Justen, F. Weigand, G. Schütz, J. Langer, D. Schmitz, H. Maletta, and R. Mattheis, Phys. Rev. B **65**, 020405(R) (2001).

[26] G. S. Chang, Y. P. Lee, J. Y. Rhee, J. Lee, K. Jeong, and C. N. Whang, Phys. Rev. Lett. **87**, 067208 (2001).

[27] P. Gambardella, S. Rusponi, M. Veronese, S. S. Dhesi, C. Grazioli, A. Dallmeyer, I. Cabria, R. Zeller, P. H. Dederichs, K. Kern, C. Carbone, and H. Brune, Science **300**, 1130 (2003).

Figure Captions

FIG. 1 (color online). (a)−(c) Perpendicular (full circles) and in-plane (open circles) hysteresis loops of the Pt/Co/Pt sandwich with a very small Pt overlayer thickness $t^{Pt}$ = 10 Å at three temperatures $T$ = 5, 180, and 300 K. (d)−(g) The perpendicular (full circles) and in-plane (open circles) hysteresis loops of the Pt/Co/Pt sandwich with a thick Pt overlayer thickness $t^{Pt}$ = 500 Å.

FIG. 2 (color online). (a) Dependence of perpendicular remanent ratio $M_r^{\perp}/M_S$ on $T$ for various $t^{Pt}$. The inset shows the dependence of the critical temperature $T_{cri}$ on $t^{Pt}$, and the solid line indicates the fitting to the data after Eq. (1). (b) Dependence of the perpendicular coercive field $H_C^{\perp}$ on $T$ for various $t^{Pt}$. The inset shows $T$ dependence of $H_C^{\perp}$ for $t^{Pt}$ = 10 Å. The solid lines are guides to the eye.

FIG. 3 (color online). Temperature dependence of the magnetization $M_S$ of the films for various $t^{Pt}$. The solid lines are guides to the eye.

FIG. 1.

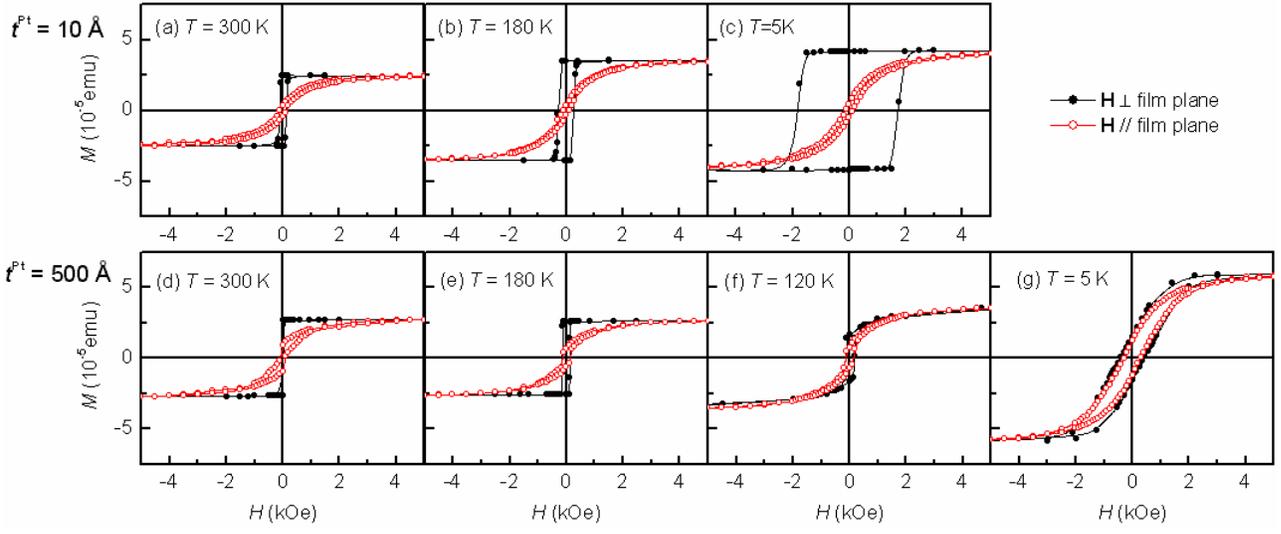

FIG. 2.

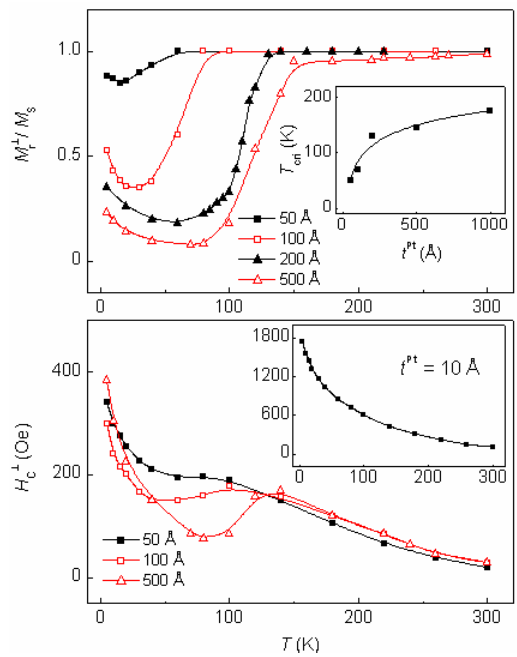

FIG. 3.

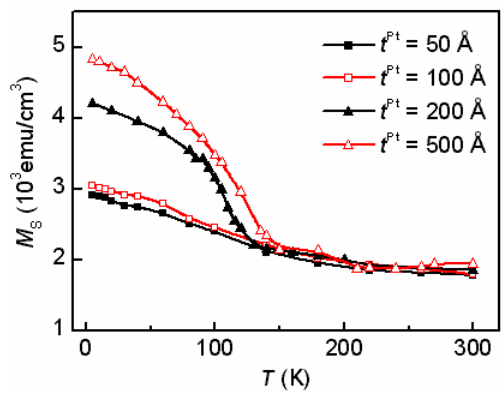